\documentstyle[aps,psfig,pre,epsf,floats]{revtex}
\begin{document}
\def\theequation{\arabic{section}.\arabic{equation}}
\global\firstfigfalse

\newcommand{\beq}{\begin{equation}}
\newcommand{\eeq}{\end{equation}}
\newcommand{\eqna}{\begin{eqnarray}}
\newcommand{\eqne}{\end{eqnarray}}
\newcommand{\dia}{\begin{displaymath}}
\newcommand{\die}{\end{displaymath}}
\newcommand{\eqnaa}{\begin{eqnarray*}}
\newcommand{\eqnae}{\end{eqnarray*}}
\def\dleft{\rlap{{\it D}}\raise 8pt\hbox{$\scriptscriptstyle\Leftarrow$}}
\def\prop{\propto}
\def\dright{\rlap{{\it D}}
\raise 8pt\hbox{$\scriptscriptstyle\Rightarrow$}}
\def\lrartop#1{#1\llap{
\raise 8pt\hbox{$\scriptscriptstyle\leftrightarrow$}}}
\def\_#1{_{\scriptscriptstyle #1}}
\def\^#1{^{\scriptscriptstyle #1}}
\def\sss{\scriptscriptstyle}
\def\dij{\delta_{\sss ij}}
\def\gij{g_{\sss ij}}
\def\Gij{g^{\sss ij}}
\def\gkm{g_{\sss km}}
\def\Gkm{g^{\sss km}}
\def\cd#1{{}_{\sss;#1}}
\def\ud#1{{}_{\sss,#1}}
\def\upcd#1{{}_{\sss;}{}^{\sss #1}}
\def\upud#1{{}_{\sss,}{}^{\sss #1}}
\def\rs{\rho^*}
\def\ro{r\_{0}}
\def\vro{{\bf r}\_{0}}
\def\qo{q\_{1}}
\def\qt{q\_{2}}
\def\lfd{L\^{D}_f}
\def\rar{\rightarrow}
\def\deriv#1#2{{d#1\over d#2}}
\def\secderiv#1#2{{d^2#1\over d#2^2}}
\def\oot{{1\over 2}}
\def\pdline#1#2{\partial#1/\partial#2}
\def\pdd#1#2#3{{\partial^2#1\over\partial#2\partial#3}}
\def\av#1{\langle#1\rangle}
\def\avlar#1{\big\langle#1\big\rangle}
\def\div{{\vec\nabla}\cdot}
\def\grad{{\vec\nabla}}
\def\curl{{\vec\nabla}\times}
\def\ma{w}
\def\DD{{\cal D}}
\def\KK{{\cal K}}
\def\FF{{\cal F}}
\def\LL{{\cal L}}
\def\MMs{{\cal M}}
\def\MM{\lrartop{\MMs}}
\def\PPs{{\cal P}}
\def\VV{{\cal V}}
\def\UU{{\vec{\bf\cal U}}}
\def\vd{\VV\_{D}}
\def\GG{{\cal G}}
\def\GGq{\GG_q}
\def\PP{\lrartop{\PPs}}
\def\AAs{{\cal A}}
\def\AA{\lrartop{\AAs}}
\def\SS{{\cal S}}
\def\BB{{\cal B}}
\def\EPS{{\cal E}}
\def\gf{\grad\f}
\def\rgf{{\bf r}\cdot\grad\f}
\def\bs{{\bf s}}
\def\bd{{\bf d}}
\def\be{{\bf e}}
\def\bff{{\bf f}}
\def\bF{{\bf F}}
\def\br{{\bf r}}
\def\bx{{\bf x}}
\def\bR{{\bf R}}
\def\bJ{{\bf J}}
\def\bV{{\bf V}}
\def\ba{{\bf a}}
\def\bA{{\bf A}}
\def\bg{{\bf g}}
\def\bn{{\bf n}}
\def\vh{{\bf h}}
\def\vu{{\bf u}}
\def\va{{\bf a}}
\def\vb{{\bf b}}
\def\dva{\d{\bf a}}
\def\vv{{\bf v}}
\def\ve{{\bf e}}
\def\vq{{\bf q}}
\def\vx{{\bf x}}
\def\vA{{\bf A}}
\def\bI{{\bf I}}
\def\vI{\bI}
\def\gfds{\grad\f\cdot \bd\bs}
\def\hm{\hat\mu}
\def\hFF{\hat\FF}
\def\eegg{\be\otimes\be\grad\grad}
\def\inv{\int\_{V}}
\def\invv{\int\_{v}}
\def\ins{\int\_{\Sigma}}
\def\inss{\int\_{\sigma}}
\def\vi{v_i}
\def\sf{S_f}
\def\si{S_i}
\def\ds{\bd\bs}
\def\dtr{d\^3r}
\def\dDr{~d\^{D}r}
\def\dDx{~d\^{D}x}
\def\dDh{~d\^{D}h}
\def\dtv{d\^3v}
\def\f{\varphi}
\def\hf{\hat\f}
\def\fl{\varphi\_{L}}
\def\gfl{\grad\fl}
\def\fx{\f\_{x}}
\def\b{\beta}
\def\z{\zeta}
\def\d{\delta}
\def\bo{\beta\_{0}}
\def\c{\gamma}
\def\co{\gamma_0}
\def\t{\tau}
\def\s{\sigma}
\def\r{\rho}
\def\rb{\r\_{B}}
\def\l{\lambda}
\def\m{\mu}
\def\n{\nu}
\def\a{\alpha}
\def\abs#1{\vert #1\vert}
\def\abgf{\abs{\gf}}
\def\abgfl{\abs{\gfl}}
\def\eps{\epsilon}
\def\pd#1#2{{\partial#1 \over \partial#2}}
\def\ad{\a\_{D}}
\def\apj{Astrophys. J.}
\def\bk{\par\noindent}
\def\vr{\br}
\def\vx{\bx}
\def\vR{\bR}
\def\vf{{\bf f}}
\def\vg{\bg}
\def\vF{\bF}
\def\vn{\bn}
\def\mi{m\_{i}}
\def\qi{q\_{i}}
\def\gfs{(\gf)\^{2}}
\def\hnmo{\hat\n^{-1}}
\def\E{E(\vr_1,...,\vr\_{N})}
\def\El#1{E(#1\vr_1,...,#1\vr\_{N})}
\def\p{\partial}
\def\Fmn{F_{\m\n}}
\def\FMN{F^{\m\n}}
\def\Jm{J^{\m}}
\def\Am{A_{\m}}
\def\e{a_0}
\def\es{\e^2}
\def\hg{\hat G}
\def\gz{\grad\z}

\title{FORCES IN NONLINEAR MEDIA}
\author{Mordehai Milgrom}
\address{Department of Condensed Matter Physics,
Weizmann Institute, Rehovot Israel}
\maketitle
\begin{abstract}
 I investigate the properties of forces on bodies in theories
governed by the generalized Poisson equation
$\div[\m(\abgf/\e)\gf]\propto  G\r$, for the potential $\f$ produced by
a distribution of sources $\r$. This
equation describes, {\it inter alia}, media with a response coefficient, $\m$,
 that depends on the field strength, such as in nonlinear, dielectric, or
 diamagnetic, media; nonlinear transport problems with
field-strength dependent conductivity or diffusion coefficient; nonlinear 
electrostatics, as in the Born-Infeld theory;
 certain stationary potential flows in compressible
 fluids, in which case the forces act on sources or obstacles in the flow.
 The expressions for the force on a point charge
 is derived exactly for the limits of very low and very high charge.
The force on an arbitrary body
 in an external field of asymptotically constant 
 gradient, $-\vg_0$, is shown to be $\vF=Q\vg_0$, where $Q$ is
 the total effective
charge of the body. The corollary $Q=0 \Rightarrow \vF=0$ is a generalization
 of d'Aembert's paradox. I show that for $G>0$ (as in Newtonian gravity)
two point charges of the same (opposite) sign still attract (repel).
The opposite is true for $G<0$. I discuss
the generalization of this to extended bodies, and derive virial relations.

\vskip 9pt
PACS numbers:
\end{abstract}


\section{INTRODUCTION}
\setcounter{equation}{0}
\par
The Poisson equation, which governs so many physical processes, has the 
nonlinear generalization
\beq \div[\m(\abgf/\e)\gf]=\ad G\r, \label{i} \eeq
by which the source distribution $\r(\vr)$, in $D$-dimensional Euclidean
space, gives rise to a potential field $\f$.
Here,
 $\e$ is a constant with the dimensions of $\gf$,
 $\ad=2(\pi)\^{D/2}/\Gamma(D/2)$
 is the $D$-dimensional complete solid angle,
 introduced here for convenience, and $G$ is a coupling constant.
 As I will show, for $ G>0$, a point, test charge is attracted to a 
(finite) point
 charge of the same sign (as in gravity), while for $ G<0$ it is repelled.
\par
 Equation(\ref{i}) describes a variety of physical problems;
some examples are:
\par
(i) Nonlinear dielectric, and diamagnetic, media; $\m$ is then the
dielectric or diamagnetic coefficient, which depends on the field
strength (here $G<0$). 
\par
(ii) Problems of nonlinear electric-current flows in systems with
 field-dependent conductivity  (nonlinear
current-voltage relation), and nonlinear diffusion problems;
 $\m(\ma)$ is the transport coefficient.
\par
 (iii) Stationary, subsonic, potential-flow problems of non-viscous fluids
with a barotropic equation of state $p=p(\varrho)$
($p$ is the pressure, $\varrho$ the density).
The stationary Euler equation is integrated into Bernoulli's equation
 $f(\varrho)=-{1\over 2}u^2 + const.$, where
$f'(\varrho)=\varrho^{-1}p'(\varrho)=c^2(\varrho)/\varrho$,
 with $c$ the speed of sound; $f$ is thus increasing with $\varrho$, and 
$\varrho$ is a function of $\abs{\vu}=\abs{\gf}$.
The stationary continuity equation then gives
$ \div[\varrho(\abgf)\gf]=s(\vr)$,
with $s$ the source density (see e.g. \cite{gt} for the ideal-gas
 case). This is eq.(\ref{i}) with $\m=\varrho$,
and $G=\ad^{-1}>0$.
 For example, in a fluid with an equation of state of the form
 $p=a \varrho^{\c}$ ($a>0,~\c\ge 1$), we have
 $\varrho(u)=\varrho(0)
 [1-(u/u_0)^2]^{1/(\c-1)}$, with $u_0^2\equiv 2c_0^2/(\c-1)$, and 
$c_0$ is the speed of sound at $u=0$. Subsonicity requires $(u/u_0)^2<(\c-1)/(\c+1)$.
\par
If we, formally, consider a stationary flow problem in a medium
with negative compressibility, $c^2<0$, ellipticity is maintained for
any value of $\gf$. For example, for a medium with a constant
$c^2<0$, $\varrho(u)=\varrho(0)exp(u^2/2\abs{c}^2)$.  
\par
(iv) Nonlinear (vacuum) electrostatics as formulated e.g. in the
Born-Infeld nonlinear electromagnetism, which also appears in
 effective Lagrangians resulting from string theory
 (see review and references in \cite{gibra}\cite{gib}).
In the original, electrostatic Born-Infeld theory
 $\m(\ma)\propto (1-\ma^2)^{-1/2}$, and $G<0$.

\par
(v) A formulation of an alternative nonrelativistic gravity to
replace the dark-matter hypothesis in galactic systems \cite{bm}.
Here $\m(\ma)\approx \ma$
 for $\ma\ll 1$, and $\m\approx 1$ for $\ma\gg 1$ ($G>0$).
\par
(vi) Equation(\ref{i}) was used in \cite{adl} as an effective-action
approximation to Abelianized QCD.
\par
(vii) Area (volume) minimization problems, 
such as the determination of the shape of a soap film
 with a dictated boundary (see e.g.
 \cite{gt}): If
 $x\_{D+1}=\f(x_1,...,x\_{D})$
 describes a $D$-dimensional hypersurface
embedded in $D+1$ dimensional Euclidean space with Cartesian coordinates
$x_1,...,x\_{D+1}$, the volume element on the surface is
$dv=[1+(\gf)^2]^{1/2}~d^Dr$. Then, eq.(\ref{i}) describes the
 problem of the minimization of the volume of the surface.
 The sources may describe a force density on the
hypersurface acting in the direction $x\_{D+1}$.
  In this problem $\m(\ma)\propto (1+\ma^2)^{-1/2}$, and $G>0$.
Born-Infeld electrostatics is the same as the area-extremization
problem for a time surface embedded in Minkowski space-time.
\par
 Much has been said in the mathematical literature on properties of the
potentials that solve eq.(\ref{i}) (see e.g. \cite{gt}).
 But, to my knowledge, very
 little has been said about forces on bodies in such theories.
 The forces can be written as
 certain integrals of $\f$, and are of obvious relevance in the
physics context.
\par
Many of the familiar and intuitive properties of the linear theory
are lost in the nonlinear case because the potential,
 and forces, is not the 
sum of the contributions of the sub-systems. For example, 
the force on a point charge
is no more proportional to the charge, does not reverse direction
when the charge of the body reverses sign, etc..
 And, Earnshaw's theorem \cite{earnshaw} no longer holds. 
Under some circumstances it is possible to
 suspend stably static charged bodies in a static
 field. This last aspect is treated in detail in \cite{sus}.
\par
The choice $\m(\ma)=\ma^{D-2}$ is special in that
the theory is then conformally invariant, and lands itself to many analytical
developments. This case is described in detail in
 \cite{conf}.

\par
After discussing  some general aspects of eq.(\ref{i}) in section II,
I take up the main subject concerning the properties of forces on bodies
in the nonlinear theories:  general properties in sections III, and 
forces on point charges in section IV.  
I conclude in section V with some examples of applications.


\section{GENERAL PROPERTIES}
\setcounter{equation}{0}

\subsection{Preliminaries}
\par
The field equation (\ref{i}) is
derivable from the action functional
\beq S=\si+\sf\equiv -\inv~\r\f\dDr
-{\e^2\over 2\ad G}\inv\FF[\gfs/\e^2]\dDr.
\label{action} \eeq 
The function $\m(\ma)$ in eq.(\ref{i}) is given by
\beq \m(\ma)=\deriv{\FF(y)}{y},~~y=\ma^2. \label{mu} \eeq
\par
For equilibrium problems, such as (i), (iv), (v), and (vii) above,
 the quantity $E\equiv -S$ may be identified with 
the energy of a charge
configuration $\r$ when $S$ is finite. I shall only be interested in 
differences in the energy between configurations with the same total
charge, and this is  finite, in general,
 even if the expression for $E$ diverges.
In the case of non-equilibrium, 
stationary transport problems, such as nonlinear diffusion,
 heat transfer, etc., $E$ is not an energy but is related to the
 entropy-generation rate.
\par
 For convenience,
 the free, additive constant in $\FF$ is chosen such that $\FF(0)=0$.
This makes the field-action density vanish at $\vr\rar \infty$ when
$\gf\rar 0$ there, and improves convergence.
I exclude from discussion theories for which  $\FF$ diverges at zero argument. I also assume $\m(\ma)>0$
 except, possibly, at $\ma=0$ where $\m$ may vanish. Thus $\FF(y)$ is an
increasing function, and is positive for $y>0$. (I assume that 
$\FF'(y)$ does not vanish for $y>0$;
 it then has a uniform sign which can always
be taken as positive by adjusting the sign of $G$.)
\par
In the area-minimization case $\FF(y)=(1+y)^{1/2}-1$.
In the flow problem $\FF(y)$ is essentially the equation of state since
the pressure is given by
$ p=p(u=0)-{u_0^2\over 2}\FF[u^2(\varrho)/u_0^2],$
with the additive constant chosen so that $\FF(0)=0$ [$p\le p(0)]$.
\par
All that I say below is straightforwardly generalized to the case where
 $\FF$,
and thus $\m$, depends explicitly on $\vr$. For instance,
in the flow problem this 
happens when the fluid is coupled to some potential field $\psi$
via $\varrho(\vr)\psi(\vr)$ for which the Bernoulli equation becomes
 $f(\varrho)=-{1\over 2}u^2-\psi(\vr) + const.$.
To avoid encumbrance I assume all along, unless otherwise said,
 that there is no such explicit
$\vr$ dependence.

\par
When the charges are not held fixed but move under the influence of
the $\f$ field, and their dynamics is of interest,
we add to the action the kinetic term for the charges
\beq S_p=\oot\int\dDr \varrho_mv^2(\vr),   \label{sp}\eeq
where $\varrho_m$ is the mass density of the particles contributing
$\r$ to the charge density.
Extremizing $S+S_p$, with respect to
 particle coordinates give the usual Euler equation
 $\varrho_m \dot{\vv}=-\r\gf$.
\par
 The solution of eq.(\ref{i})  is unique inside
a volume $V$ when one dictates on its boundary the value of $\f$ or that of
 $\m(\abgf/\e)\p_n\f$ (or a combination thereof) 
($\p_n\f$ is the normal component of $\f$)  provided the function $\n(\ma)\equiv\ma\m(\ma)$ is
 an increasing function (see e.g. \cite{sol} for a  proof).
This monotonicity of $\n(\ma)$--which I shall assume all along--is
 tantamount to $\hm\equiv dln~\m(\ma)dln~\ma > -1$.
 This is also the condition for
the ellipticity of eq.(\ref{feqa}). In the case of a stationary flow we have
$\hm = -u^2/c^2$; the ellipticity condition
 is then equivalent to subsonicity of the flow.
\par
 I now show that from
 the ellipticity condition and the choice $\FF(0)=0$ follows that
 the logarithmic derivative of $\FF$,
 $\hFF(y)\equiv y\FF'(y)/\FF(y)$, satisfies
  $\hFF(y)>1/2$ for all $y>0$ for which ellipticity obtains.
  I use this inequality repeatedly in what follows. Define
\beq \chi(\ma)\equiv \FF(\ma^2)[2\hFF(\ma^2)-1]= 2\m(\ma)\ma^2-\FF(\ma^2),
\label{opareta} \eeq
 (where $y=\ma^2$).
First note that $\chi'(\ma)=2\ma\m(1+\hm)$,
so $\chi'>0$ for $\ma>0$.
Since $\chi(0)=0$, $\chi(\ma)>0$ for $\ma>0$.
Thus $\hFF(y)>1/2$ for all $y>0$ [as $\FF(y)>0$ for $y>0$].
(The condition $\FF(0)=0$, on which
the derivation of $\hFF>1/2$ depends, does indeed enter in the cases where
this inequality is used below.) 
\par
It is useful to write our theory for a general curved space whose metric is
$\gij$ ($\Gij$ its inverse, and $g=\abs{det(\gij)}$). 
The covariant form of the action is then
\beq S=-\inv~g^{1/2}\r^*\f\dDr
-{\e^2\over 2\ad G}\inv g^{1/2}\FF[(\f\upud{i}\f\ud{i})/\e^2]\dDr,
\label{covaction} \eeq
with $\f\upud{i}=\Gij\f\ud{j}$ ($\f\ud{i}\equiv \p\f/\p x^i$),
and $\r^*\equiv g^{-1/2}\r$ is a scalar under general coordinate
transformations. Repeated indices are summed over.
 The covariant form of the field equation is
\beq [\m(\f\upud{i}\f\ud{i}/\e^2)\f\upud{k}]\cd{k}=\ad G\r^*,
\label{cogatiza} \eeq
where a semi-colon signifies a covariant derivative. The covariant
divergence appearing in eq.(\ref{cogatiza}) is given in terms of
the normal divergence of a vector $v^k$ as 
$v^k\cd{k}=g^{-1/2}(g^{1/2}v^k)\ud{k}$.
 So, using usual derivatives instead we have
\beq  [g^{1/2}\m(\f\upud{i}\f\ud{i}/\e^2)\f\upud{k}]\ud{k}=\ad G\r.
\label{cogapipa} \eeq

\par
From the covariant action we can derive the 
 the field stress tensor (the energy-momentum tensor
when working in Lorentzian space-time).
This is the functional derivative of the
field action  with respect to the metric:
 under a variation $\d \gij$
\beq \d \sf=\oot\int g^{1/2}\d\Gij\PPs_{ij}\dDr.  \label{laplata} \eeq
In the Euclidean case, on which I concentrate hereafter, we find
\beq \PP=-{\e^2\over 2\ad G}(\FF-2\m\gf\otimes\gf)=
-{\e^2\over 2\ad G}\FF(1-2\hFF\ve\otimes\ve),
 \label{prussa} \eeq
where $\ve\equiv \gf/\abs{\gf}$ is a unit vector along $\gf$.
 The trace of $\PP$ is
$-(\e^2/ 2\ad G)\FF(D-2\hFF)$.
The field direction $\ve$ is an eigenvector of $\PP$ with eigenvalue 
$-(\e^2/ 2\ad G)\FF(1-2\hFF)$.
The inequality $\hFF>1/2$ 
 implies that this eigenvalue is always positive:
 there is always tension along the field lines. All other
 eigenvalues are equal, and negative.
\par
For solutions of the field equation the divergence of $\PPs_{ij}$,
which measures the rate of change of the momentum density, is given by
\beq \div\PP=\r\gf. \label{kuopa} \eeq
This conservation law can be derived directly from the Euclidean action
(\ref{action}) and follows from
 its translation invariance: under infinitesimal translations
$\vr\rar\vr+\va$, $\f(\vr)\rar\f(\vr)+(\va\cdot\grad)\f$, etc.
$S\rar S+\va\cdot(\div\PP-\r\gf)$.
\par
 The field equation can also be written as
\beq \m\AAs\_{ij}\f\ud{i}\ud{j}=\ad G\r, \label{feqa} \eeq
 where
 \beq  \AA=  (1+\hat\m \ve\otimes\ve), \label{mata}\eeq
 with $\hm$ the
logarithmic derivative of $\m$, and $\ve\otimes\ve$
is the metrix whose $(i,j)$ element is $e_ie_j$
 (all dependent on $\gf$).
 Since $\hm>-1$ $\AA$ is positive definite.
If $\m(0)=0$, points where $\gf=0$ need special treatment which I do not
 go into here. 
\par
If we make a small change $\d\r$ in $\r$, the field equation can be 
linearized in the small change $\z$ in $\f$ to read \cite{sol}
\beq \div[\m\AA\cdot\grad\z]=\ad G\d\r, \label{ipofur} \eeq
This is the same as the
 equation for the electrostatic potential produced by the density
$\d\r$ in a linear dielectric medium with a position-dependent, anisotropic
dielectric constant $\m\AA$. 
\par
  A variation of $\f$ in the volume $V$ gives rise to the
a variation in $S$
\beq \d S={1\over \ad G}\inv \d\f\{\div[\m(\abgf/\e)\gf]-\ad G\r\}\dDr
-{1\over\ad G}\ins\m~\d\f\gf\cdot\ds, \label{var} \eeq
where $\Sigma$ is the boundary of $V$.
For  potentials that solve the 
field equation, and hence nullify the first term in eq.(\ref{var}),
 we have 
\beq \d S=-{1\over\ad G}\ins\m~\d\f\gf\cdot\ds. \label{vara} \eeq
We can obtain useful integral constraints on such solutions
--such
as conservation laws and virial
relations--by considering specific variations that do not
nullify the surface term.
 Some examples are given in Appendix A.
\par
The second-order change in the action (energy) is
\beq \d^2 E=-\d^2 S={1\over 2\ad G}\inv \m\grad\d\f\cdot\AA\cdot\grad
\d\f\dDr.   \label{metrefa} \eeq
 The ellipticity condition makes the integral
 positive (when $\gf\not\equiv 0$), and thus a
solution of the field equation is a minimum of the energy for $ G>0$,
and a maximum for $ G<0$.
\par
Using the integral relation (\ref{viria}) derived in Appendix A,
 which holds for solutions of the
 field equation, to eliminate the explicit dependence of $E$ on the sources
we get
\beq E=-S=
-{\e^2\over 2\ad G}\inv\FF[\gfs/\e^2](2\hFF-1)\dDr.
\label{amtabun} \eeq
So, in light of the above inequality for $\hFF$, $E$ is positive for
$G<0$ (as in electrostatics).
\par
The field equation enjoys a certain scaling property\cite{sol} in that
if $\f(\vr)$ and $\r(\vr)$ are a consistent pair then so is
\beq \f_{\l}(\vr)=\l\f(\l\^{-1}\vr),~~~
\r_{\l}(\vr)=\l\^{-1}\r(\l\^{-1}\vr), \label{vuiops} \eeq
 with the appropriately scaled boundary conditions. Charges then scale as
$q_{\l}=\l\^{D-1}q$.


\subsection{Asymptotic behavior of the potential}
\par
When the medium can be considered infinite, a common choice of boundary
condition, describing an isolated system, is $\gf\rar 0$ at infinity.
 In fact, if the potential can be assumed to become
spherical at infinity, this boundary requirement sometimes 
follows from potential
equation itself through Gauss's theorem. 
 The behavior of $\m(\ma)$
near $\ma=0$ is then relevant. For concreteness,
  I assume in the rest of the paper that $\FF$, and thus $\m$,
 approaches  a power of the argument near 0.
\beq \m(\ma)\rar \ma\^{\bo},~~~~\FF(y)\rar {2\over 2+\bo}
y\^{(2+\bo)/2}~~~~(\bo>-1).  \label{bet} \eeq
This choice does not cover the case in \cite{adl}, where $\m$ is
logarithmic at small values of the argument.
\par
If the sources $\r$
 are contained within a finite volume, and the total charge, $Q$, does
not vanish, the field becomes
radial at infinity, and, applying Gauss's theorem to the field equation
for a sphere of a large radius $r$, we find asymptotically
\beq \gf\approx s(QG)\abs{\hg}^{\co}\abs{Q}^{\co}
r\^{-\co(D-1)}\vn. \label{asym} \eeq
Here
 $\hg\equiv  G\e^{\bo}$, $\co\equiv 1/(1+\bo)$,
 $s(a)=sign(a)$, and $\vn\equiv \vr/\abs{\vr}$.
\par
The value $\bo=D-2$ is a limiting one. For higher values of $\bo$,
$\f$ diverges like a power of $r$ at large $r$ (when $Q\not =0$)
with $\gf$ still vanishing there.
 For the limiting case,
$\f$ is logarithmic at infinity,
 and $\gf\propto \vr r\^{-2}$. In this case the
action is infinite when calculated for the whole space.
 The revised theory of
Newtonian gravity discussed in \cite{bm} is of this limiting-power type,
and so are the class of conformally invariant theories with $\hm=D-2$
discussed in \cite{conf}.
In what follows I assume $\bo\le D-2$ unless otherwise stated (as in the
 discussion of one-dimensional systems).
\par
For $Q=0$ I do not have a general expression for the asymptotic behavior of
the field. Because of the nonlinearity, multipoles are no more very relevant.
For  $\bo=D-2$ the asymptotic form can be obtained as follows.
In the exact-power-law theory with $\b=D-2$ one can use the conformal 
invariance. Take the origin at a point outside charges, and make a 
conformal transformation  $\vr\rar a^2\vr/r^2$, where $a$ is the radius of the 
reflection sphere. If $\f(\vr)$ is the solution of the original problem
then $\hat\f(\vr)\equiv\f(a^2\vr/r^2)$ is the solution of the new problem
with the transformed charge distribution
 $\hat\r(\vr)=(a/r)^{2D}\r(a^2\vr/r^2)$,
 which also has a vanishing total charge (see
\cite{conf} for more details).
So the asymptotic behavior of $\f$ is obtained from the behavior of
$\hat\f$ at the origin (where the charge density is 0). I assume that
 $\hat\f$ is analytic there, so its dominant behavior
is, generically, $\hat\f\approx  a^{-2}{\bf K}\cdot\vr$,
 where ${\bf K}$ is some constant vector.
 This gives for the generic, asymptotic behavior of $\f$
\beq  \f\approx {\bf K}\cdot\vr/r^2. \label{asymom} \eeq
${\bf K}$ might be viewed as the 
asymptotic-behavior dipole, but it is not proportional to the dipole of the 
charge distribution. If ${\bf K}=0$ the asymptotic behavior is, more
 generally, of the form 
\beq  \f\approx K_{i\_{1}...i\_{n}}
r_{i\_{1}}...r_{i\_{n}}/r^{2n}, \label{asypap} \eeq
where $a^{-2n}K_{i\_{1}...i\_{n}}$, a totally symmetric constant tensor,
 is the first non-vanishing Taylor coefficient
in the expansion of $\hat\f$ at the origin. (I am discussing only the 
leading behavior, not expansion terms. Because of the nonlinearity
the higher-order behavior depends on the leading one.)
 Basically, I use the conformal
invariance to argue that $\f$ has to be analytic in $\vr/r^2$ at infinity, and 
hence to greatly constrain its form there.
The Ks have further to satisfy
algebraic relations insuring that expression(\ref{asypap}) satisfies
the vacuum field equation. For n=1 there are no extra relations on  ${\bf K}$.
For $n=2$ the algebraic relation 
 in the linear ($D=2$) case is the usual tracelessness requirement
 $\sum_i K_{ii}=0$. For $D>2$ the condition on the matrix $\KK$, the elements
 of which are $K_{ij}$, is
$(\vr\KK\vr)^{(D-4)/2}\vr[(D-2)\KK^3+Trace(\KK)\KK^2]\vr=0$ for all
 vectors $\vr$.
This, for the symmetric $\KK$, can be shown to imply
 $\KK=0$, so there is 
no dominant $n=2$ behavior for $D>2$. 
Now consider a general theory with $\bo=D-2$.
 Let $\f$ be the solution 
for a confined charge distribution with vanishing total charge.
Define
\beq \r^*\equiv(\ad G)^{-1}\div[\bar\m(\abgf/\e)\gf], \label{iopitar} \eeq
where $\bar\m$ is the exact $D-2$ power. Because asymptotically (where 
$\gf\rar 0$) we have $\m\rar\bar\m$, we get from Gauss's theorem that $\r^*$
also has a vanishing total charge. And, if $\m$
 approaches its power-low behavior
fast enough, $\r^*$ will be well bounded. 
But from eq.(\ref{iopitar}) $\f$ is a
 solution for $\r^*$ in the theory with the exact power-law $\bar\m$ and so,
from the arguments above, 
must also have the asymptotic form(\ref{asymom})(\ref{asypap}).
\par
 Outside a spherical distribution of zero total charge
the field vanishes for all theories.

\par
Also of interest is the boundary condition $\gf\rar-\vg_0$ for $\vr\rar\infty$
($\vg_0$ is a constant vector).
 It pertains e.g. to a system of charges
 in an external electric field, to magnetized
or superconducting bodies in an external magnetic field, or
 to obstacles and sources in an asymptotically uniform flow.
For the asymptotic field
use eq.(\ref{ipofur}) to linearize in
 $\gz\equiv\gf+\vg_0$:
\beq  \div[\gz+\hm\ve_0(\ve_0\cdot\gz)]=0,  \label{utinama} \eeq
 and solve with
 $\gz\rar0$ at infinity
($\ve_0\equiv\vg_0/\abs{\vg_0}$).
Taking $\ve_0$ to be the $x_1$-direction we see that, asymptotically,
 $\z$ satisfies the
Laplace equation in the coordinates
 $\vr'\equiv[(1+\hm)^{-1/2}x_1,x_2,...,x\_{D}])$.
Thus, $\z$ has the standard multipolar asymptotic expansion in $\vr'$.
When the total charge, $Q$, is finite the dominant behavior
is $\z\propto Q/r'^{D-2}$ ($D>2$).


\section{FORCES ON BODIES}
\setcounter{equation}{0}
\par
A body is an isolated region, $v$, where the field is externally
 disturbed in one way or
 another. For example, a body may be defined by some rigid distribution of
 charges
 in $v$, or by dictating the potential or its gradient on the boundary of 
$v$. The body 
may also be defined as a ``$\m$ inclusion'': dictating in $v$ a  $\m(\ma)$
that is different form the ambient one.
 Examples of bodies defined by boundary conditions are:
 a rigid body in the flow problem and
 a superconducting body in a magnetic field, for both of which the normal
 component of $\gf$ vanishes at the surface; 
 a conducting (equipotential) surface
in an electric field; and a restricting boundary in the
 volume-minimization problem, for which the potential is dictated.
Examples of bodies defined by $\m$ inclusions
 are a dielectric inclusion,
 or, in the flow problem, a region of space where
a fluid is subject to an external potential $\psi(\vr)$ that couples to the 
fluid density. There is then an extra term, $\psi(\vr)$,
  in the Bernoulli equation,
 and $\varrho(u^2)$ is then also a function of
$\vr$ in $v$ through $\psi$.
\par
 Consider then a background 
having some ambient, position-independent $\m$.
Within this background are 
 embedded disjoint bodies of different types as defined
 above. Once the solution, $\f$, it gotten for the system
 one can replace the bodies by effective
charge distributions
 that give the same potential field outside the bodies. The effective charge
 density is simply
\beq \rs(\vr)\equiv (\ad G)^{-1} \div[\m(\abgf/\e)\gf] \label{ilbuz},\eeq
with the ambient $\m$ used everywhere. If a body is defined by boundary
conditions we continue the solution inside the body by solving the
field equation with the ambient $\mu$ and imposing the same boundary 
conditions for the internal solution.
Obviously, the effective charges given in this way appears as an extra charge
 only on the 
bodies because, by the field equation, the right-hand side of 
expression(\ref{ilbuz}) gives the true charges outside bodies. For example, a 
body that is defined by dictating $\f$ on its surface is replaced by
a  $\rs$ that constitutes a surface
 charge; a body defined by dictating $\m\p_n\f$ on the surface is replaced 
 by a surface dipole layer (whose total charge obviously vanishes).
A body defined by a $\m$ inclusion is replaced by an
effective charge distribution whose total effective charge is the same as the 
actual charge on the body. This is because Gauss's theorem, applied
to eq.(\ref{ilbuz}), gives the total effective 
charge as a surface integral over $\m\gf$ on a surface
surrounding the body, but, from the field equation, this also equals the 
actual total charge of the body. So the total effective charge of a
$\m$ inclusion always vanishes 
if there are no actual sources within it.
\par
The force on a body can be defined in several equivalent ways. For example,
it may be taken as the gradient of the total energy under translation of 
the body: 
First replace all bodies in the system by their effective charges. Then
 translate the body in question, rigidly and 
infinitesimally, by $\dva$, keeping all the other charges (including effective
ones) fixed. For the energy increment $\d E=-\dva\cdot\vF_v$, we identify
$\vF_v$ as the force on the body.
 Under the translation, the effective charge distribution, $\r$, changes by
$\d\rs=-\dva\cdot\grad\rs$ in $v$, and $\d \rs=0$ outside.
Since $\f$ is an extremum of $S$, the change
in $E$ ($=-S$) can be calculated as if only $\rs$ has changed 
($\d\f=0$ at infinity). Thus
\beq \dva\cdot\vF_v=-\d E=-\invv~\f\d\rs\dDr=\dva\cdot\invv~\grad\rs~\f\dDr=
-\dva\cdot\invv\rs\gf\dDr,   \label{pjutara} \eeq
where the last equality is obtained by integrating by parts.
Thus,
\beq \vF_v=-\invv\rs\gf\dDr=\invv\f\grad\rs\dDr.   \label{pjonura} \eeq
The first equality may also serve as a definition of the force: it says that
 the force is the sum of forces on the elements of the body, each given by
$-\rs\gf\dDr$. But, unlike the linear case, here $\f$ cannot be taken as 
that due to the rest of the system (excluding the body from the sources).
In the linear case, but not here, the potential can be written as the 
sum of the contribution of the body and of the rest of the system. Since
a body does not exert a force on itself (this is also true in the nonlinear
 case, see below) the contribution of the self field
 drops from the expression for the force.
\par
A third way--perhaps the most useful--to write the force uses 
eq.(\ref{kuopa}) ($\rs\gf=\div\PP$) and
expression (\ref{prussa}) for $\PP$ to get 
\beq \vF_v=-\inss\PP\cdot\ds={\e^2\over 2 \ad G}[\inss\FF~\ds
-\inss 2\FF\hFF\abgf^{-2}\gf\gf\cdot\ds].  \label{forcea} \eeq
The integration is done over any closed
 surface, $\sigma$, that surrounds the body
 and excludes all other sources and bodies
 (compare with the expression of the force as a surface
 integral in \cite{bm}). This expression has the advantage that it
does not require the effective charge distribution of the body, and
 employs only the field outside the body.
\par
The surface integral in eq.(\ref{forcea}) vanishes automatically
 for the surface at infinity in theories with $\bo\le D-2$,
 so the total force on a whole system vanishes,
 as expected from translational invariance. In other theories (e.g. in one
dimension--see below)
 the boundary conditions for an isolated body  must insure this.
\par
For nonlinear dielectrics, diamagnetics, etc.
 the above definition of the force coincides with the usual expression
for the electromagnetic force.
In the flow problem $\vF_v$ is the mechanical force acting on a region
containing sources ($\r$ replaced by $s$) because the integral in
 expression (\ref{pjonura}) ($\int s\vu$) is 
the rate at which the sources in the volume impart momentum to the
flow. For a rigid obstacle standing in the flow, expression (\ref{forcea})
gives the mechanical force the flow exerts 
on the obstacle because the second integral vanishes
as $\gf\cdot\ds$ vanishes on the surface (the flow is parallel to the
surface) and we are left with
 $\vF_v=-\int p\ds$.
The same is true for a type-I superconducting body in a nonlinear
magnetized medium: due to the Meissner effect $\gf\cdot\ds=0$ on the
surface of the body.
\par
In the volume-minimization problem the $\vF$ is
the actual lateral force
 (in the $x_1,...,x\_{D}$ plane) on the volume in question due to the uneven
tension.
\par
In the case of non-equilibrium, stationary-transport problems, such
 as nonlinear diffusion,
the action functional is not an energy, and $\vF$ is not a force, but measures
the gradient in the entropy-generation rate when sources are 
translated.

\subsection{A body in an external field--extension of the d'Alembert paradox}
\par
We saw that when $\gf$ vanishes at infinity the total force on any bound
system vanishes. When we have
 $\gf\rar -\vg_0$ at infinity, describing a constant external field
in which the system is immersed, it can be shown that even in the 
nonlinear case the force on the whole 
system is  $Q\vg_0$, where $Q$ is the total charge. This can be seen by
using the asymptotic form of the field given below eq.(\ref{utinama})
in expression (\ref{forcea}) for $\vF$ on a surface going to infinity.
In the linear case this result is trivial since the total force
is the sum of the total mutual forces of the charges, which vanishes, and the
sum of the external forces $\int \r\vg_0~\dDr=Q\vg_0$.
It follows from this, for example, that the force on an arbitrary
$\m$ inclusion in a constant external field vanishes since, as we
 saw above, its total effective charge vanishes.
The application of this result to the compressible flow problem
constitute a generalization of the well-known d'Alembert paradox
in fluid mechanics to the effect that 
a single, static obstacle in a non-viscous, incompressible,  potential
 flow that is
 uniform at infinity is subject to no force.
We now see, more generally, that the paradox applies as well to the 
nonlinear (compressible but subsonic) case, and to
 any single body made of fluid sources (or sinks), rigid obstacles, regions
where body forces apply, or any other configuration that can be replaced
by an effective charge distribution with vanishing total charge.


\subsection{Point bodies}
\par
When a body is very small compared with the typical scale of the
system  we can approximate it by a point charge (PC)
$q$ at position $\vro$, with density $\r(\vr)=q\d\^{D}(\vr-\vro)$.
We may
 view the body as the limit of some finite-size charge
\beq \r(\vr)=q\lim_{\l\rar 0} \l\^{-D}\hat\r(\vr/\l), \label{gusta} \eeq
where $\hat\r$ is some smooth, finite charge 
distribution normalized to
$\int\hat\r(\vR)d^DR=1$. 
Not all problems admit point charges. Applying Gauss's theorem to
eq.(\ref{i}) we see that $\ma\m(\ma)$ diverges as $r^{-(D-1)}$
near a point charge, so such charges are not admitted in theories
with $\ma\m(\ma)$ bound from above. This is the case for the 
volume-minimization problem, where $\ma\m(\ma)<1$. It is also the case
 in the flow problem with $c^2>0$, where subsonicity, and hence
ellipticity, is lost near a point source.
For ``flow'' problems with $c^2<0$, ellipticity is maintained
at all values of $\gf$, and point sources are not objectionable.
\par
The concept of a PC is useful only if the field
everywhere in a system containing a PC
is independent of the particular choice of the structure
function $\hat\r$ of the PC, so that it
 enters only though its total charge. I was not able to prove that
this always holds. One possible route to proving this is to show
 that  an infinitesimal change
$\d\hat\r$ in the structure function, which does not change the total charge,
 $\int\d\hat\r(\vr)\dDr=0$,
 produces an everywhere-vanishing increment
 in the  potential.
 The potential increment $\z$  is a solution of the linear equation 
(\ref{ipofur}) (where the background field depends on $\l$)
with $\d\r=\l^{-D}\d\hat\r(\vr/\l)$, in the limit $\l\rar 0$.
 In this limit all the moments of $\d\r$
vanish, which might tell us that $\z$ vanishes:
Start with the charge distribution
  $\r_{\l\l'}=\l^{-D}\hat\r(\vr/\l)+\l'^{-D}\d\hat\r(\vr/\l')$ 
in place of the point PC.
 Let $\f_\l$ be the solution of the problem for $\d\r=0$,
and $\f_\l+\z_{\l\l'}$ the solution for $\r_{\l\l'}$.
In the limit $\l\rar 0$,  $\f_\l$ goes to  the solution for the PC
having $\hat\r$ as a structure function. In the limit $\l=\l'\rar 0$,
$\f_\l+\z_{\l\l'}$ goes to the required solution for a PC with
structure function $\hat\r+\d\hat\r$.
 We want to show that in this last limit $\z_{\l\l'}\rar 0$.
Instead of going to the limit $\l=\l'=0$ along the $\l=\l'$ line we first 
take the limit $\l'\rar 0$ for finite $\l$ and then take the limit $\l\rar 0$.
$\z_{\l\l'}$ solves eq.(\ref{ipofur}) with the left-hand side depending on
 $\l$, and the right-hand side source being $\l'^{-D}\d\hat\r(\vr/\l')$. 
As all the moments of this source vanish in the limit $\l'\rar 0$, I conclude
 that $\z_{\l\l'}\rar 0$ in the limit. Taking now the limit $\l\rar 0$
we are left with $\z=0$ in the limit. The remaining loophole concerns the
equality of the two limits. 
\par
 Be this as it may, the results concerning PCs are only valid
when the field does not depend on the structure function. 
Employing the expression for the force as a surface integral
eq.(\ref{forcea}) we see that the force on a PC is then also
independent of the choice of $\hat\r$.

\par
Zero-size bodies with higher multipoles can, of course, also be constructed
 by considering a multiple limit with point charges of infinite
charges at zero distances. Such bodies with pure multipole charge
distribution do not have the special role they have in the
linear case. They do not, in general,
produce unique-multipole fields, and, unlike the pure-charge case
the force on them depends on details of
their structure,
not only on the components of the multipole.

\subsection{Test bodies}
\par
Consider a sub-system of charge distribution  $\rb$.
 In the limit $\rb\rar 0$ it can be considered as a test body:
its contribution to the
full potential can be neglected in expression (\ref{pjonura})
for the force on itself, and we have
$\vF=-\int\rb\grad\hf$, where $\hf$ is the potential determined by the
rest of the system. A point charge has an infinite density. But, it may still be
considered a test charge (the force on which is $\vF=-q\grad\hf$) provided
$q$ is small enough that a surface can be drawn around it
such that (1) the surface is small compared with the scale over which
 the field varies appreciably, (2) it is far enough from the
 charge that the latter's contribution on the surface
 is small compared with the field due to the rest of the system alone
(which is then approximately constant on the surface).


\subsection{Attraction or repulsion?}

\par
 I now show that it is still correct in the nonlinear case that 
 two equal point charges attract, and two opposite
 charges repel each other (for $ G>0$; and vice versa for $G<0$).
 Take,  more generally, two finite, disjoint  bodies with
charge distributions that are the mirror images of each other
about some $(D-1)$-dimensional
 hyperplane. The force on each can be calculated by taking
the symmetry hyperplane as the integration surface in eq.(\ref{forcea})
(the integration on the hemisphere at infinity vanishes with our choice
$\FF(0)=0$).
Since $\f$ is symmetric about the hyperplane, $\gf$ is in the
plane. We thus get from
 eq.(\ref{forcea})
\beq  \vF_v={\e^2\over 2\ad G}\inss\FF~ \ds,
 \label{forlata} \eeq
which is attractive for $G>0$.
 If one body is the negative-charge reflection of the
other, $\f$ is antisymmetric,
 and $\gf$ is perpendicular
to the hyperplane so that $\PP\cdot\ds\propto \FF(1-2\hFF)\ds$, and
from eq.(\ref{forcea})
\beq  \vF_v={\e^2\over 2\ad G}\inss\FF(1-2\hFF) \ds,
 \label{forbana} \eeq
which is repulsive for $G>0$ since $\hFF>1/2$ when $\gf\not= 0$.
\par
Consider, more generally, two disjoint bodies $B_i$ 
defined by charge distributions $\r_i>0$ in the non-overlapping volumes
 $v_i$, i=1,2.
It is meaningful to ask whether they attract or repel each other only if
 they can be separated by some $(D-1)$-dimensional hyperplane. 
We have attraction if
 the force on $B_1$ crosses  any such separating hyperplane
 from the side of $B_1$ to that of $B_2$
(see Fig. 1).
\begin{figure}
\centerline{\psfig{figure=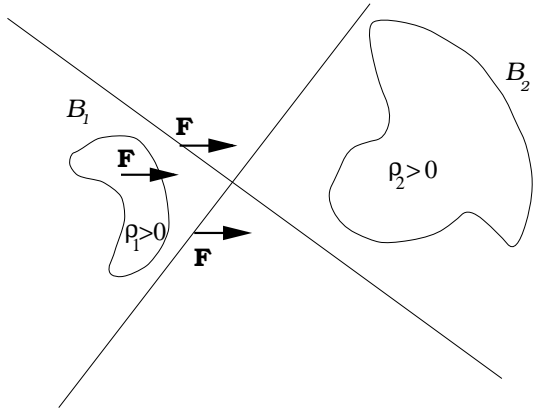,width=7cm}}
\caption{The force $\vF$ on body 1 crosses every separating plane from 
the side of 1 to side of 2}
\end{figure}
 I conjecture without a general proof that indeed
this is always the case, and that if the two bodies are oppositely
charged, i.e. if, say, $\r_2<0$, they always repel each other.
(The sign of the charge within each body must be uniform.)
It follows from the result proved in Appendix C
 that the conjecture holds when one of the bodies, say $B_1$, is
spherically symmetric with a density profile decreasing 
from the center out (I thank Shoshana Kamin for discussions leading
 to this proof). In fact, in this case the statement is stronger: for any
 hyperplane $H^*$ through the center of $B_1$ with $B_2$ wholly
to its one side (side 2) the force on $B_1$ points from side 1 to side 2
(see Fig. 2).
\begin{figure}
\centerline{\psfig{figure=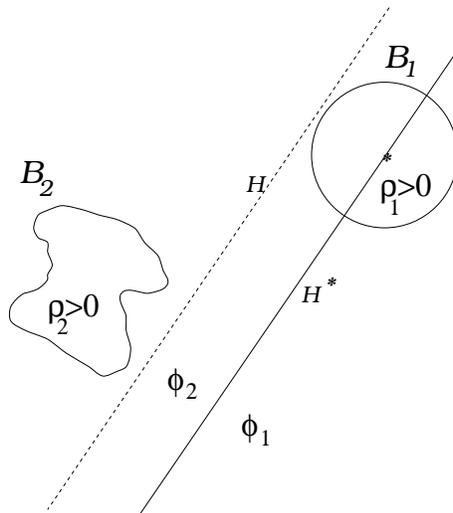,width=6cm}}
\caption{The setup for demonstrating attraction of like charges
when one of the bodies is spherical with a decreasing density
profile}
\end{figure}

\par
 We learn from the above result that the force on a 
spherical body with a decreasing density profile is always within
the convex closure of the cone defined by its center and the  other body
(the envelope of all the planes through the point that are
 tangent to the body).
\par
Because a point charge may be considered as 
the limit of such a spherical body, we deduce that the 
force on a point charge in the presence of an extended body of
uniform-sign charge is always within the cone from the point to the
 convex closure of the body. In particular, two point charges of the same sign
 always attract each other.
\par
All the above generalizes, {\it mutatis mutandis}, to oppositely charged
 bodies.

\par
\begin{figure}
\centerline{\psfig{figure=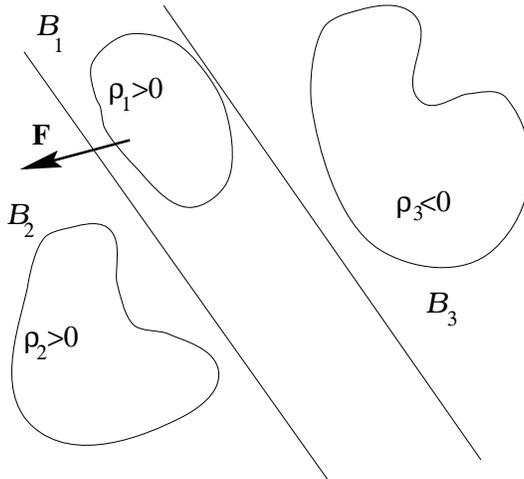,width=7cm}}
\caption{A configuration depicting the push-pull conjecture:
for any pair of parallel separating planes the force $\vF$ on $B_1$
crosses the planes into side 2}
\end{figure}
The two body case may be generalized to a push-pull conjecture concerning 
three bodies (as in Fig. 3) with 
$\r_1,~\r_2>0, ~\r_3<0$: for every
 two parallel hyperplanes separating the three bodies
 the force $\vF$ acting on $B_1$ crosses the planes into the side of $B_1$.
This seems to be a good way to summarize
the concept of attraction-repulsion of bodies with constant-sign charges as
all else regarding the question follows from it.
\par
I have not been able to prove this push-pull conjecture in all
generality. It is proved in Appendix C when body 1 is 
spherically symmetric with radially decreasing density; so, in particular,
the  conjecture holds when $B_1$ is a point charge.
 The conjecture is also proved in the one-dimensional case (see below).
It also holds, generally,  when body 1 is a collection of test
 charges. It then follows from the comparison principle discussed
in Appendix B.
 In this case  the force on body 1 is $\vF=-\int_1\r\gf_{12}$, where
 $\f_{12}$ is the potential produced by bodies 2 and 3 alone. The comparison
principle tell us that on any plane 
separating these bodies $-\gf_{12}$ crosses from the side of 3 to that of 2
so $\vF_1$ does so as well.
 Similarly, the conjecture holds when bodies 2 and 3
are made of test particles. The force on body 1
can then be written as $\vF=\int_{12}\r\gf_{1}$, where the integration 
is performed
over bodies 1 and 2, and $\f_1$ is produced by body 1 alone. By
 the corollary of the comparison principle, 
 $\gf_{1}$ crosses from the side of 3 to that of 2 in the volume of body
 2 , and the opposite in the volume of body 3. Thus,  the integrand, 
and hence $\vF$, crosses from 3 to 2.

\par
Take a system containing  charges of a uniform sign, say $\r>0$, with $\S$ a
 convex surface surrounding all the charges. It follows that on $S$, $\gf$
points  outward (inward when $\r<0$).
 So, for example, in theories where $\m\le 1$
we have for the total charge, $Q$, the inequality
$Q=(\ad G)^{-1}\ins\m\gf\cdot\ds \le (\ad G)^{-1}\ins\gf\cdot\ds\equiv Q^*$.
$Q^*$ is the charge that would be deduced from the field $\gf$ in a linear 
theory.


\subsection{Forces in the one-dimensional case}
\par
The one-dimensional case can be solved in closed form once the boundary 
conditions are fixed. If we require that $\gf(\infty)=-\gf(-\infty)$ so as to
nullify the force on an isolated system, then $\gf\rar const.$ at $\infty$.
The force, $F$, on
 an arbitrary charged body
of total charge $q$, in the presence of a charge distribution
that does not overlap with it depends only on $q$ and on
 the difference, $Q$, between the total charges to the right and
 to the left of the body.
\beq F=F(q,Q)={\e^2\over 8 G}[\chi(\ma^+)-\chi(\ma^-)],\label{lateqa} \eeq
where $\chi(\ma)$ is defined in eq.(\ref{opareta}),
and
\beq \ma^{\pm}=s( G)\n^{-1}\left({\abs{ G}\over 2\e}
\abs{q\pm Q}\right).  \label{katareca} \eeq

We see that $F$ is invariant under translations of the body, as long
as it does not cross other charges, and that
\beq F(-q,Q)=-F(q,Q)=F(q,-Q),~~~F(Q,q)=F(q,Q).\label{jata}\eeq
These are peculiarities of the one-dimensional case. In higher dimensions
the magnitude of the force on a charge
 does not, in general, remain invariant when the sign of
 the charge is reversed (all others kept intact).
\par
Since $\chi'>0$ for $\ma>0$, $\chi$ is increasing everywhere. Thus,
 $F$ does not vanish unless $\ma^+=\ma^-$,
i.e., unless $q=0$, or $Q=0$.
The push-pull conjecture hold here as is easily seen.


\section{ FORCES ON POINT CHARGES}
\setcounter{equation}{0}

\subsection{The limit of a very large point charge}
\par
 Consider a charge distribution $\r$ and a
 point charge $q$ that does not overlap with it. Assume also that $q$
 is much larger than any partial charge that makes $\r$ up
(i.e. $\int_{v}\r\dDr\ll q$ for any volume $v$). 
 We can then consider $\r$ to be a collection of test charges relative to $q$.
 From momentum
 conservation, the force $F_q(\vR)$ acting
 on $q$ at position $\vR$ is opposite the force $F_{\r}$ acting
 on the distribution $\r$. This latter is given in the test-particles limit
 by $F_{\r}=-\int\dDr \r(\vr)\gf_q$, where $\f_q$ is
the field produced by $q$ alone, which is straightforwardly gotten through
 Gauss's theorem to give
\beq \vF_q(\vR)=s(q G)\e
\int\dDr\r(\vr)\n^{-1}\left({\abs{q G}\over \e \abs{\vr-\vR}\^{D-1}}
\right){\vr-\vR\over\abs{\vr-\vR}}, \label{lutaru} \eeq
 with $\n(\ma)\equiv \ma\m(\ma)$.
This force is derivable from an effective potential $E_q$:
\beq E_q(\vR)=\int\dDr\r(\vr)\GGq(\abs{\vr-\vR}). \label{gratar} \eeq
where the effective Green's function, $\GGq$, satisfies
\beq \grad_{\vr}\GGq(\vr)=s(q G)\e
\n^{-1}(z)\vn,  \label{gutareq} \eeq
with $z\equiv\abs{q G}/ \e \abs{\vr}\^{D-1}$, and $\vn={\vr\over \abs{\vr}}$.
\par
The effective potential is then a linear functional of the
density $\r$. It satisfies the Poisson equation
\beq \Delta E_q(\vR)=\int\dDr\r(\vr)
\Delta\GGq(\abs{\vr-\vR}), \label{brazar} \eeq
 with
\beq \Delta \GGq(\vr)=s(q G)(D-1)\e
\abs{\vr}^{-1}{\ma\hm(\ma)\over 1+\hm(\ma)},  \label{qurates} \eeq
where $\ma=\n^{-1}(z)$.
\par
The above results apply also to any finite spherical body of very large
total charge, and are straightforwardly 
 extended to the case of an arbitrary body whose
$\f$ field is known.

\subsection{The two-body force}
\par
For two point charges $\qo\le\qt$, a distance $\ell$ apart,
the vanishing of the total force and moment
 tell us that the forces
on the two charges are opposite each other, and lie along the connecting line;
write its magnitude   $f(\qo,\qt,\ell)$.
We have seen above that $f$ has the sign of $G\qo\qt$
 ($f$ is positive for attraction). Because $\ell$ is the only length scale,
 we can reduce
the number of variables to two independent, dimensionless variables
 constructed from   $\qo,\qt,\ell$, and $\e$; for example,
 $-1\le\eta\equiv\qo/\qt\le 1$ and
$z\equiv \abs{G\qt}/\e\ell\^{D-1}$.
 Since $\e\qo$ has dimensions of force,
we can write, for example, 
\beq f=s(G\qt)\e\qo\hat f(\eta,z). \label{tofju}\eeq
\par
Our earlier discussion implies some constraints of $\hat f$.
 For instance, when one charge is much smaller than the other
 $\abs{\eta}\ll 1$, the test-charge result tells us that
\beq \hat f(\abs{\eta}\ll 1,z)= \n^{-1}(z), \label{boozsa} \eeq
to lowest order in $\eta$ [$z=\n(\ma)=\ma\m(\ma)$].
\par
 For a power-law medium with $\m(\ma)=\ma^\b$ one deduces from scaling
 properties of the field equation that the $\ell$-dependence of the force 
is $\ell^{-\c(D-1)}$, where  $\c=1/(1+\b)$. So here
\beq \hat f=\z(\eta)z^\c. \label{layura} \eeq
From eq.(\ref{boozsa}) $\z(0)=1$.
Since, in general, $\z(-1)\not=\z(1)$, the forces for equal and for
opposite charges are
not of the same magnitude, in contradistinction with the linear case.
For the special case $\b=D-2$, the two-body force was found in closed 
form\cite{conf}
\beq f(\qo,\qt,\ell)=s(G){1\over \ell}
d\^{-1}\abs{G\e^\b}^{d-1}(\abs{\qo+\qt}^d
-\abs{\qo}^d-\abs{\qt}^d)  \label{forcof} \eeq
[$d\equiv D/(D-1)=1+\c$].
(The forces in a three-point-charge system of zero total charge were also
 derived in \cite{conf}.)
For two equal charges $\qo=\qt=q$,
 $f=2s(G)\ell^{-1}d^{-1}\abs{\hg}^{d-1}\abs{q}^d(2^{d-1}-1)$, while for
 opposite
charges: $\qo=-\qt=q$,
 $f=-2s(G)\ell^{-1}d^{-1}\abs{\hg}^{d-1}\abs{q}^d$. The two are equal in
magnitude only in the (linear) two-dimensional case. Interestingly, in the
 limit of large dimension, where $d\rar 1$, the two-body force for two charges
of the same sign vanishes as $D^{-1}$ [from eq.(\ref{forcof})],
 while the force for opposite charges does not.
This can be generalized: for a given configuration of
$N$ point charges {\it of the same sign} the force on each becomes smaller as
$D^{-1}$ in the limit of large $D$  (letting also the theory's power increase,
 $\b=D-2$).
\section{ SOME EXAMPLES OF APPLICATIONS}
\setcounter{equation}{0}
\par
I next discuss a number of potential applications, some of which I
alluded to earlier.
I concentrate on two of the physical problems listed in the
 introduction, {\it viz} stationary (potential) flows of barotropic
compressible fluids, and media with field dependent dielectric, or
 diamagnetic, constants.
\par
Concerning the flow problem, we saw that the d'Alembert paradox
can be extended to the compressible case. So, 
the force acting on a body of total effective charge in a flow of
constant asymptotic speed vanishes. This implies the vanishing of
the force on a body of rigid walls standing in such flow, on a system
made of sinks and sources with vanishing total out-flux.
If the fluid is electrically charged--weakly, so that the fluid
does not self-interact--then the flow is modified
by the presence of a region with an electric field. The effect
of such a region can be described by an effective
 sink-source distribution whose total out-flux vanishes. So, the 
net force on such a region due to the fluid motion vanishes.
We also learned from the discussion of the sign of the forces
that in such flows two sinks, or two sources always attract each 
other while a source and a sink repel each other. 
\par
For the nonlinear dielectric, the generalized d'Alembert  paradox
says that, as in the linear case, the force on an arbitrary charge
 distribution, $e_i$, of vanishing net charge 
in a constant external field, $\vec E$,
vanishes, even though the force on an individual component of 
charge is not $e_i\vec E$. From this we learn that,
for example, the force on a dielectric inclusion, or an
 equipotential body of zero net charge, vanishes in a constant
external field. We also deduce that the force vanishes 
on a superconducting inclusion
in a nonlinear magnetic medium in a constant magnetic field.

\vskip 5pt
\begin{acknowledgements}
I thank Shoshana Kamin for helpful suggestions.
\end{acknowledgements}



\setcounter{section}{0}
\def\thesection{\Alph{section}}
\def\theequation{\Alph{section}.\arabic{equation}}


\section{Integral relations}
\def\thesection{\Alph{section}}
\def\theequation{\Alph{section}.\arabic{equation}}
\setcounter{equation}{0}
\par
\vskip 5pt
We employ the technique described in \cite{vir} to derive certain
useful virial relations directly from the action by
substituting in eq.(\ref{vara}) for $\d S$ various choices of $\d\f$.
 If either $\bo<D-2$, or $\bo=D-2$ and $Q=0$, 
 the potential vanishes at infinity. We can then obtain one
 relation by taking $\d\f=\eps\f$, with $\eps$ infinitesimal.
The vanishing of $\f$ at infinity leads to the vanishing of $\d S$.
But $\d S$  can also be calculated directly to yield a virial relation
\beq \inv~\r\f\dDr+{1\over \ad G}\inv\gfs\m(\abgf/\e) \dDr
=0. \label{viria} \eeq
(This can also be derived by multiplying $\f$ by the expression for
$\r$ from the field equation, and integrating by parts.)
\par
In the relation above (and below) $\r$ should be understood to include
true sources as well as all the effective sources replacing boundary
conditions on closed surfaces in the system.
\par
Another useful relation, satisfied by solutions of the field equation,
may be obtained by considering variations of $S$
 produced by dilations of space
coordinates: $\f(\vr)\rar\f(\l\vr)$. For an infinitesimal
dilation $\l=1+\eps$, we have $\d\f=\eps\vr\cdot\gf$.
 Now, $\d S$ can also be
calculated directly by substituting
 $\f(\vr)\rar\f(\l\vr)$ in $S$;
 then taking the derivative with respect to
$\l$ at $\l=1$ to obtain for the finite volume
\beq \left({dS\over d\l}\right)_{\l=1}=
 -\inv\r\vr\cdot\gf\dDr+{\es\over 2\ad G}\inv[D\FF
-2{\gfs\over\es}\m]\dDr
-{\es\over 2\ad G}\ins\FF\vr\cdot\ds. \label{surfb}\eeq
So, comparing with expression(\ref{vara}) for the variation, we get
 an expression for the virial $\VV$
in the volume $V$.
\beq\VV\equiv\inv\r\vr\cdot\gf\dDr={\es\over 2\ad G}\inv\FF(D-2\hFF)\dDr
+\ins\vr\cdot\PP\cdot\ds.  \label{virbb}\eeq
This can also be obtained by substituting in the definition of $\VV$
the expression for $\r$ from the field equation, and integration by
 parts.
The value of $\VV$ is independent of the choice of origin due to
momentum conservation.
\par
Taking the surface to infinity, the surface term in eq.(\ref{virbb})
can be evaluated. At this point I restrict myself to $D\ge2$;
the one-dimensional case is exactly solvable and is treated below.
 For $\bo<D-2$ the surface integral vanishes at infinity, and
\beq  \VV=
{\es\over 2\ad G}\int\FF(D-2\hFF)\dDr.
\label{virco}\eeq
 For $\bo=D-2$ the surface integral at infinity converges to yield
\beq \VV={\es\over 2\ad G}\int\FF(D-2\hFF)\dDr
+d^{-1}\hg\^{-1}\abs{\hg Q}^d,  \label{surfc} \eeq
where $d\equiv D/(D-1)$, and $\hg\equiv G\e^{\bo}$.
The conformally invariant theories discussed in \cite{conf}
  have $\hFF=D/2$ and so
 $\VV=d^{-1}\hg\^{-1}\abs{\hg Q}^d.$
This has been put to extensive use in \cite{conf}.
\par
Another integral of interest is
\beq \UU\equiv \inv\r[\vr(\vr\cdot\gf)
-\oot r^2\gf]\dDr.\label{pjulara}\eeq
By substituting $\r$ from the field equation and integrating by parts
we arrive at the expression
\beq \UU={\es\over 2\ad G}\inv\FF(D-2\hFF)\vr\dDr
-{\es\over 4\ad G}\ins\FF r^2\{(1-2\vn\otimes\vn)+
\hFF[4(\vn\cdot\ve)\vn\otimes\ve-2\ve\otimes\ve]\}\cdot\ds
  \label{gavutira}\eeq
($\vn=\vr/\abs{\vr}$, $\ve=\gf/\abs{\gf}$).
\par
Note that $\UU$ depends, in general, on the choice of origin.
Using the vanishing of the total force and total moment we see that if
 the origin is shifted by $-\va$, $\UU$ is changed by $\VV\va$.


\def\thesection{\Alph{section}}
\def\theequation{\Alph{section}.\arabic{equation}}

\section{ THE COMPARISON PRINCIPLE AND SOME CONSEQUENCES}
\setcounter{equation}{0}
\par
First I show that if $\f_1,~\f_2$ are continuous functions that 
solve our equation for densities $\r_1\ge\r_2$
 in a volume $V$ of boundary $\Sigma$, and $\f_1\le\f_2$ on $\Sigma$, then
$\f_1\le\f_2$ everywhere in $V$. This is known  as a comparison
principle for solution of elliptic equations (e.g.\cite{gt}). I give here a 
proof that applies specifically to the form of our equation and is thus more
elementary than the proofs found in the literature.
Start with the identity
\beq \int_v(\f_1-\f_2)(\r_1-\r_2)\dDr\propto
 \int_v(\f_1-\f_2)\div(\m_1\gf_1-\m_2\gf_2)\dDr= \eeq

\beq \int_{\sigma}(\f_1-\f_2)(\m_1\gf_1-\m_2\gf_2)\cdot\ds
-\int_v(\gf_1-\gf_2)\cdot(\m_1\gf_1-\m_2\gf_2)\dDr\label{luop}\eeq
[where $\m_i=\m(\abs{\gf_i})]$,
gotten by using the expression for $\r_i$ from the field equation,
 use of Gauss's theorem, and integration by parts.
I want to show that the region $v$ in $V$, in which $\f_1>\f_2$ is
empty. If there is even one point where $\f_1>\f_2$, then, by continuity
of $\f_1-\f_2$, $v$ must contain
a whole (non-zero-measure) neighborhood.
 Apply identity \ref{luop} to this volume $v$. On its boundary
$\sigma$ we have $\f_1=\f_2$ (from continuity of $\f_i$), whether $\sigma$
has overlap with $\Sigma$, or is completely interior to $V$.
Thus the first term on the right-hand-side of \ref{luop} vanishes.
It can be shown (see \cite{sol}) that, in light of the ellipticity
condition, the integrand in the second term is non-negative, and vanishes
only where $\gf_1=\gf_2$. This is because
 when $\ma\m(\ma)$ is non-decreasing
$(\va-\vb)\cdot [\m(\abs{\va})\va-\m(\abs{\vb})\vb]\ge 0$ for any two
vectors $\va,~\vb$, and vanishes only for $\va=\vb$.
However, by the assumptions, the left-hand side is non-negative
and so the right-hand side must vanish, and hence $\gf_1=\gf_2$
in $v$. Thus, $\f_1-\f_2$ is constant in the whole region where it is
 positive. This, however, contradicts
the assumption that $\f_1\le\f_2$ on $\Sigma$ with the continuity of the
 potentials.
\par
I now apply the theorem to the following useful
 configuration. Let  $\f(x_1,...,x\_{D})$ be the solution for a
source distribution that satisfies
 $\r(x_1,...,x\_{D})\ge\r(-x_1,...,x\_{D})$. The
 boundary condition at 
infinity is  $\f(\vr)\rar s(r)$.
 Look on the two halves of the solution in the two
half-spaces separated by the  $x_1=0$ hyperplane ($H^*$)
 as two solutions of the field equation
in a half-space $x_1\geq 0$; so, 
$\f_1(x_1,...,x\_{D})\equiv\f(x_1,...,x\_{D})$, and 
$\f_2(x_1,...,x\_{D})\equiv\f(-x_1,...,x\_{D})$.
 We have $\f_2(0,x\_{2},...,x\_{D})=\f_1(0,x\_{2},...,x\_{D})$.
So, $\f_1$ and $\f_2$ are solutions with the same
 boundary values (they also have the same boundary condition at $\infty$).
The comparison principle then tells us that
 $\f_1(x_1,...,x\_{D})\le\f_2(x_1,...,x\_{D})$, or
$\f(x_1,...,x\_{D})\le\f(-x_1,...,x\_{D})$.
In particular, $\pd{\f}{x_1}(x_1=0)\le 0$. $H^*$ can be any plane
separating a source distribution $\r_1\ge 0$ from $\r_2\le 0$.
\par
Let $\r>0$ in a volume $v$ be the whole the source distribution,
 and $A$ its convex 
closure ($A$ is the smallest convex volume containing all the points where
$\r\not = 0$). Then the gradient of the potential at any point $\vr$
 outside $A$ points away from $A$ (because it points away from the side of 
$A$ on any plane separating $\vr$ from $A$).
\par
We can deduce from this that
for a system of two point charges of the same sign
 the tangent to the field lines alway crosses the
line connecting the charges between the two. For oppositely charges points
the tangent crosses outside this line segment. This generalizes the usual 
situation familiar from the linear case where it follows simply from the
 vector addition of the two forces due to the point charges.


\section{PROOF OF THE PUSH-PULL CONJECTURE FOR A
 SPHERICAL BODY}
\setcounter{equation}{0}
\par
Consider a system of three disjoint bodies: $B_1$, which is
 spherically symmetric with $\r_1>0$ decreasing from the center out;
 $B_2$ with $\r_2>0$; and  $B_3$ with $\r_3<0$. $H^*$ is a plane
through the center of $B_1$ that separates $B_2$ from $B_3$.
I show that the force $\vF$ on $B_1$ crosses $H^*$ from side 3 to side
2. The push-pull conjecture for this special case
 is a weaker statement and follows as a corollary. 
\par
Choose the coordinates such that $H^*$ is the $x_1=0$ plane, with $B_2$ on the
$x_1<0$ side. We then have 
from the previous appendix that 
$\f(-x_1,...,x\_{D})\leq \f(x_1,...,x\_{D})$ for $x_1>0$. 
Using the second expression in eq.(\ref{pjonura}) to calculate the $x_1$
component of the force on 
$B_1$, and employing the latter's symmetry, we have
\beq F_1=\int_{x_1>0}\dDr~[\f(x_1,...,x\_{D})-\f(-x_1,...,x\_{D})]\p_{x_1}\r
\leq 0. \label{luftare} \eeq
where use was made of the fact that $\p_{x_1}\r\leq 0$
for $x_1>0$, as $\r$ is monotonically decreasing with radius.

\end{document}